\documentclass[preprint2]{aastex}

\usepackage{graphicx}
\usepackage{amsmath}

\begin{document}

\title{Measurement of the lowest millimetre-wave transition frequency of the CH radical}

\author{S. Truppe, R. J. Hendricks, E. A. Hinds and M. R. Tarbutt}
\affil{Centre for Cold Matter, Blackett Laboratory, Imperial College London, Prince Consort Road, London, SW72AZ, UK.}
\email{m.tarbutt@imperial.ac.uk}


\begin{abstract}
The CH radical offers a sensitive way to test the hypothesis that fundamental constants measured on earth may differ from those observed in other parts of the universe. The starting point for such a comparison is to have  accurate laboratory frequencies. Here we measure the frequency of the lowest millimetre-wave transition of CH, near 535\,GHz, with an accuracy of 0.6\,kHz. This improves the uncertainty by roughly two orders of magnitude over previous determinations and opens the way for sensitive new tests of varying constants.
\end{abstract}


\section{Introduction}

The CH radical is an important constituent of stellar atmospheres and interstellar gas clouds and plays an essential role in most combustion processes. It is a well established tracer for molecular hydrogen, and is a basic constituent of interstellar chemistry, being one of the building blocks for more complex species. Recently, the lowest-lying $\Lambda$-doublet transitions of CH, and the lowest rotational transition, have been identified as highly sensitive to a possible variation of the fine-structure constant, $\alpha$ or of the electron-to-proton mass ratio, $\mu$ \citep{Kozlov(1)09, Nijs(1)12}. Such variations - in time, in space, or with local matter density - are expected in some higher-dimensional theories that aim to unify gravity with the other forces, and in some theories of dark energy \citep{Uzan(1)03}. The ground state $\Lambda$-doublet transition, at 3.3\,GHz, has been detected in numerous interstellar clouds of the Milky Way \citep{Rydbeck(1)76, Hjalmarson(1)77} and in several other galaxies \citep{Whiteoak(1)80}, and has been used to study the velocity structure, physical conditions and chemistry of these clouds. The $\Lambda$-doublet transition of the first rotationally excited state, at 0.7\,GHz, has also been detected in several clouds \citep{Ziurys(1)85}. Constraints on the variation of $\alpha$ and $\mu$ with local matter density have recently been obtained using astronomical observations of the CH $\Lambda$-doublets, along with new laboratory measurements \citep{Truppe(1)13}. As pointed out by \citet{Nijs(1)12}, very sensitive and robust measurements could be obtained by observing the lowest rotational transition of CH, which has components at 532.8 and 536.8\,GHz, along with the two $\Lambda$-doublets at 3.3 and 0.7\,GHz, all from the same cloud. Since this test uses only the four lowest-lying states of a single species, it is quite robust against systematic shifts caused by spatial segregation. Using the Herschel telescope, the lowest rotational transition has recently been detected in emission from star-forming regions of the Milky Way \citep{vanderWiel(1)10}, and in absorption in interstellar clouds against the mm-wave background of star-forming regions \citep{Gerin(1)10, Qin(1)10}. Using ALMA, this transition has also now been observed in the absorber PKS 1830-211 at a red-shift of $z=0.89$ \citep{Muller(1)13}, opening up the prospect of using CH to search for variation of the constants over cosmological time, as well as with matter density. For these tests to reach the highest possible sensitivity, an improved laboratory measurement of the rotational transition frequency is needed.

\begin{figure}[t]
\includegraphics[width=0.9\linewidth]{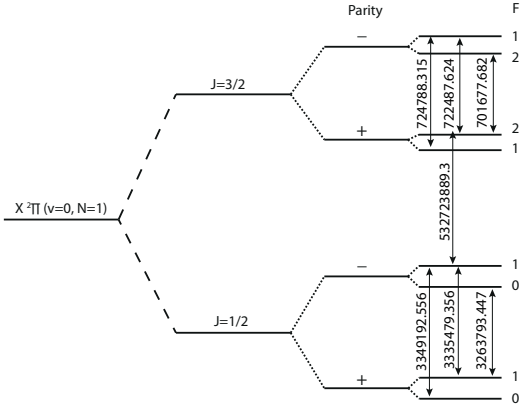}
\caption{\label{Fig:Levels} Relevant energy levels in the X$^{2}\Pi (v=0)$ ground state of CH, using Hund's case (b) notation. Transition frequencies are in kHz and are given by the present measurement in combination with the $\Lambda$-doublet transitions given in \citet{Truppe(1)13}.}
\end{figure}

Figure \ref{Fig:Levels} shows the relevant energy levels in the ground state of CH. The two lowest levels, having angular momenta $J=1/2$ and $J=3/2$, are separated by a combination of rotational and spin-orbit energy. Each of these is split into a $\Lambda$-doublet, consisting of two levels of opposite parity, $p$. The magnetic moment of the hydrogen nucleus causes a further splitting into hyperfine components labelled by $F$, the total angular momentum quantum number. We use the notation $(J^p,F)$ to label these levels. Precise measurements of the $\Lambda$-doubling frequencies are given in \citet{Truppe(1)13}. The previous most precise measurements of the $J=1/2-3/2$ transitions are given by \citet{Amano(1)00}, with uncertainties between 30 and 100\,kHz. Rotational and spin-orbit transitions between higher-lying rotational levels were measured with similar precision by \citet{Davidson(1)01}, who then used all the available data on the $v=0$ level of the X$^{2}\Pi$ state to determine a precise set of molecular parameters. Here we measure the $J=1/2-3/2$ frequencies with a precision of 0.6\,kHz.

\begin{figure}[t]
\includegraphics[width=1.0\linewidth]{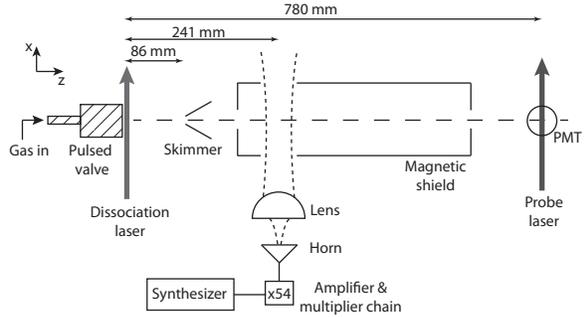}
\caption{\label{Fig:Setup} Schematic of the experiment.}
\end{figure}

\section{Experiment}

Figure \ref{Fig:Setup} shows the experimental setup. We produce a supersonic beam of CH by photo-dissociating bromoform.  At 4\,bar pressure, a carrier gas of He, Ne, Ar or Kr  is bubbled through liquid bromoform (CHBr$_3$, 96\% purity stabilized in ethanol) and then expands through the 1\,mm-diameter nozzle of a pulsed solenoid valve into a vacuum chamber \citep{Lindner(1)98, Romanzin(1)06}. Light from an excimer laser (wavelength of 248\,nm, duration 20\,ns and energy up to 220\,mJ) is focused in front of the nozzle to a spot 1\,mm high (along $y$) and $4$\,mm wide (along $z$), where it dissociates the bromoform to produce the CH. This source is pulsed with a repetition rate of 10\,Hz. At $z\!=\!86$\,mm the molecules pass through a 2\,mm-diameter skimmer into a second vacuum chamber where the pressure is below $10^{-7}$\,mbar. Here, they enter a magnetically shielded region, and then, at $z=241$\,mm, pass through a beam of mm-wave radiation which drives a selected hyperfine component of the $1/2^{-}-3/2^{+}$ transition. The radiation propagates along $x$ and is linearly polarized along $y$. After leaving the magnetic shield the molecules are detected at $z=780$\,mm by driving the $\text{A}^2\Delta (v=0) \leftarrow \text{X}^2\Pi (v=0)$ transition with a cw laser beam (from a frequency-doubled titanium-sapphire laser), and imaging the resulting fluorescence onto a photomultiplier tube. This probe light propagates along $x$, is linearly polarized along $z$, has a wavelength near 430.15\,nm, and has a power of about 5\,mW in a rectangular cross section 4\,mm high and $1.4$\,mm wide. The pulse of fluorescence is recorded with a temporal resolution of about 5\,$\mu$s. To measure the depletion of the $J=1/2$ population, or the increase in the $J=3/2$ population, the laser drives the transitions to the $J=3/2$ or $J=5/2$ levels of the $\text{A}^2\Delta (v=0,N=2)$ state respectively. The frequencies of these transitions are given by \citet{Zachwieja(1)95}, where they are designated  as R$_{\text{22ff}}(1/2)$ and R$_{\text{11ff}}(3/2)$.

The mm-wave radiation comes from an amplifier-multiplier chain (Virginia Diodes, AMC) which generates the 54th harmonic of a frequency synthesizer, phase-locked to a 10\,MHz GPS reference. A WR1.5 diagonal horn antenna couples the radiation out of the waveguide, delivering about 13\,$\mu$W at 535\,GHz. The electric field distribution at the output is approximately gaussian with a waist of 1\,mm. This beam is collimated by a plano-convex teflon lens of 30\,mm focal length, then passes into the vacuum chamber through a quartz window. At the position of the molecular beam, the electric field profile of the mm-wave beam is approximately $e^{-(y^{2}+z^{2})/w^{2}}$, where $w \simeq 5$\,mm.

\section{Results and discussion}

\begin{figure}[tb]
\includegraphics[width=0.9\linewidth]{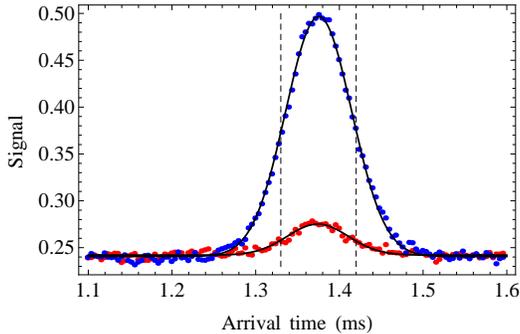}
\caption{\label{Fig:Tof} Time-of-flight profiles for molecules having $J=3/2$ before (red points) and after (blue points) driving the $(1/2^{-},1)-(3/2^{+},2)$ transition. Solid lines are gaussian fits to the data. The molecules used in the analysis are those that arrive between the dashed lines.}
\end{figure}

With Ar as the carrier gas, figure~\ref{Fig:Tof} shows the time-of-flight profile for molecules in the $J=3/2$ state, both with and without the mm-wave radiation applied. From the peak and width of these profiles, we deduce that the molecules have a mean speed of 567\,m/s and a translational temperature of 0.4\,K. When Kr is the carrier gas, the speed is 412\,m/s, and for Ne it is 791\,m/s. When the mm-wave radiation is off, the $J=3/2$ fluorescence signal is about 5\% of the $J=1/2$ signal. When the mm-wave radiation is on, and resonant with the molecular transition, the $J=3/2$ signal increases by a factor of 8.

\begin{figure}[t]
\includegraphics[width=0.9\linewidth]{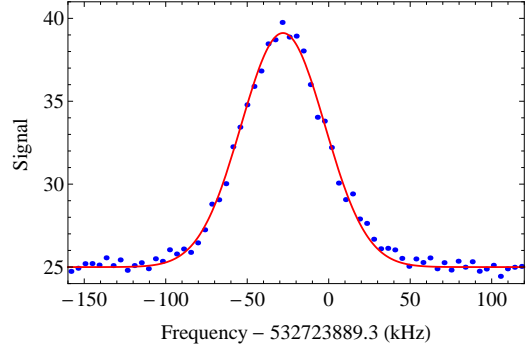}
\caption{\label{Fig:Spectrum} $J=3/2$ signal versus mm-wave frequency, showing the $(1/2^{-},1)-(3/2^{+},2)$ transition. The line is a gaussian fit to the data.}
\end{figure}

Figure \ref{Fig:Spectrum} shows how the laser-induced fluorescence signal from the $(3/2^{+})$ state depends on the mm-wave frequency, which is tuned near the $(1/2^{-},1)-(3/2^{+},2)$ transition and stepped between shots of the experiment in a random order. Each point in the spectrum is the integral of the time-of-flight profile between the limits indicated by the dashed lines in figure \ref{Fig:Tof}. To understand the lineshape, we have modelled the experiment, taking into account the gaussian intensity distribution and wavefront curvature of the mm-wave beam, and the Doppler broadening due to the range of transverse velocities that are detected. Broadening due to the Zeeman effect (see below) is negligible. This model predicts a gaussian lineshape with a full width at half maximum (FWHM) of 44\,kHz for a speed of 412\,m/s, 58\,kHz for 567\,m/s and 79\,kHz for 791\,m/s. Gaussian fits to our data give reasonably good agreement with these predictions: we find FWHMs of $51 \pm 4$\,kHz, $62 \pm 2$\,kHz, and $79 \pm 4$\,kHz for these three speeds, where the uncertainties are the standard deviations of several measurements. The fit indicated by the solid line in figure \ref{Fig:Spectrum} determines the line centre with a precision of 0.3\,kHz.

\begin{figure}[t]
\includegraphics[width=0.9\linewidth]{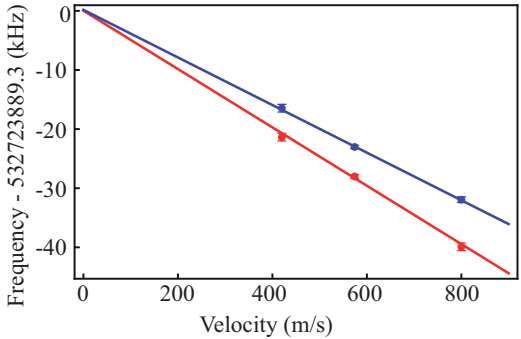}
\caption{\label{Fig:Doppler} Doppler shift of the $(1/2^{-},1)-(3/2^{+},2)$ transition frequency as a function of velocity, for two different angles of the mm-wave beam.}
\end{figure}

Inevitably, there is a Doppler shift because the mm-wave beam is not exactly perpendicular to the molecular beam. We correct for this by measuring at all three beam velocities and extrapolating the line centre to zero velocity, as plotted in figure~\ref{Fig:Doppler}.  For each velocity, we repeat the measurement a number of times and take the weighted mean of the line centres. As expected, we observe a linear shift of the frequency with velocity due to the Doppler shift. To check the reliability of this method, we change the propagation direction of the mm-wave beam and repeat the measurements, as also plotted in figure \ref{Fig:Doppler}. The two sets of data have different slopes due to the change in angle, but they have the same zero-velocity intercept, which we take as the true resonance frequency. The slopes correspond to angles of 28 and 22\,mrad with the x-axis.  The same procedure is also used to measure the frequency of the $(1/2^{-},1)-(3/2^{+},1)$ hyperfine component of the transition.

Next we consider systematic shifts due to the Zeeman effect. The g-factor is close to zero when $J=1/2$, whereas for $J=3/2$ it is 1.081 when $F=1$ and 0.648 when $F=2$. The $(1/2^{-},1)-(3/2^{+},2)$ transition is therefore split by a small magnetic field into five equally spaced components separated by 9.07\,Hz/nT. In the shielded region where the molecules interact with the mm-wave beam, we measure a magnetic field component along $z$ of 13\,nT, with the components along $x$ and $y$ being at least ten times smaller. The linear Zeeman splitting in this field is only 120\,Hz. Furthermore, this linear splitting should not shift the line centre because, for a linearly polarized mm-wave beam, the shifted components are symmetric about the line centre. We have verified this by applying a field large enough to resolve the Zeeman splitting. We have also measured the effect of the quadratic Zeeman shift due to magnetic dipole matrix elements off-diagonal in $F$. We apply magnetic fields along $z$ that are large enough to observe a shift of the line centre. As expected, this shift is quadratic, and has a curvature of -20\,Hz/$\mu$T$^{2}$. We conclude that, in our residual field, shifts due to both linear and quadratic Zeeman shifts are negligible at the current level of precision.

The dc Stark shift due to stray static electric fields, the motional Stark shift due to the motion of the molecules through the magnetic field, and the ac Stark shift due to frequency sidebands, are all negligible in the experiment. Also negligible are frequency shifts due to collisions, blackbody radiation and the second-order Doppler shift.

Our measured frequencies for the $(1/2^{-},1)-(3/2^{+},2)$ and $(1/2^{-},1)-(3/2^{+},1)$ transitions are $532723889.3 \pm 0.7$\,kHz and $532721588.5 \pm 1.4$\,kHz respectively. When taken together with the measured $J=1/2$ and $J=3/2$ $\Lambda$-doublet transition frequencies \citep{Truppe(1)13}, these results, determine the frequencies for all components of the $(N=1,J=1/2)-(N=1,J=3/2)$ transition in $^{12}$CH. These are given in Table \ref{tFinal}. The uncertainties in the table are slightly smaller than those of the individual measurements quoted above because there are more measurements than independent frequencies. Our results are between 50 and 150 times more precise than the previous best measurements \citep{Amano(1)00}, and differ from those by up to 3.6 standard deviations. Subtracting our results from those of \citet{Amano(1)00} and taking the weighted mean, we find a difference of $79 \pm 15$\,kHz, indicating an uncontrolled systematic error in those previous measurements.

\begin{table}[tb]
\centering
\begin{tabular}{c|c}
\hline\hline
Transition & Frequency (kHz) \\
\hline
$(1/2^+,0)-(3/2^-,1)$ & $536795569.5\pm 0.6$\\
$(1/2^+,1)-(3/2^-,1)$ & $536781856.3\pm 0.6$\\
$(1/2^+,1)-(3/2^-,2)$ & $536761046.3\pm 0.6$\\
$(1/2^-,0)-(3/2^+,1)$ & $532793274.6\pm 0.6$\\
$(1/2^-,1)-(3/2^+,2)$ & $532723889.3\pm 0.6$\\
$(1/2^-,1)-(3/2^+,1)$ & $532721588.6\pm 0.6$\\
\hline
\end{tabular}
\caption{Frequencies of all components of the $(N=1,J=1/2)-(N=1,J=3/2)$ transition of $^{12}$CH.}
\label{tFinal}
\end{table}

The absolute accuracy of these new laboratory frequencies is improved to almost 1 part-per-billion, at which level the uncertainty no longer hinders the search for varying fundamental constants. This improvement will also allow more accurate velocity determinations to be made in astrophysical measurements. If some need were to emerge for measuring the laboratory frequency even more accurately, this could be readily achieved using Ramsey's method \citep{Ramsey(1)50} of two excitation zones. In the present apparatus, this would increase the frequency resolution by a factor of 100 without any loss of molecular flux. The use of resonant cavities around the two excitation zones would eliminate the residual Doppler shift, and increase the amount of available power for driving the transition. With careful shielding to reduce systematic shifts due to magnetic fields \citep{Truppe(1)13}, a measurement of this THz frequency interval with an accuracy of 1\,Hz is feasible. Precise measurements of sub-millimetre-wave transition frequencies in other molecules could also be made using this method.

\acknowledgments
We thank Ben Sauer and Heather Lewandowski for useful discussions and assistance in the lab. This work was supported in the UK by the EPSRC and the Royal Society.

 \end{document}